\documentclass[two column,prb,multicol,amsmath,amssymb,showkeys]{revtex4}
\usepackage[dvips]{graphicx}
\usepackage{graphicx}% Include figure files
\usepackage{dcolumn}% Align table columns on decimal point
\usepackage{bm}% bold math
\usepackage{graphics}
\usepackage{soul}
\usepackage{epsfig,color}
\usepackage{hyperref}

\begin{document}
\title[J. Vahedi ]{Shot Noise of Charge and Spin Current of a Quantum Dot Coupled to Semiconductor Electrodes}
\author{Zahra Sartipi and Javad Vahedi$\footnote{email: javahedi@iausari.ac.ir\\ Tel: (+98)9111554504\\Fax: (+98)151 33251506}$}
\address{Department of Physics, Sari Branch, Islamic Azad University, Sari, Iran.}
\date{\today}
\begin{abstract}
Based on the scattering matrix theory and non-equilibrium green function method, we have investigated the fluctuations of charge and spin current of the systems which consists of a quantum dot (QD) with a resonant level coupled to two semiconductor contacts within in alternative site $(AS)$ and alternative bond $(AB)$ framework, where two transverse $(B_x)$ and longitudinal $(B_z)$ magnetic fields are applied to the QD. It is only necessary to use the auto-correlation function to characterize the fluctuations of charge current for a two-terminal system because of the relation which is defined as $\sum_{\alpha}^e S_{\alpha\beta}$ = $\sum_{\beta}^e S_{\alpha\beta}=0$. Our result shows that both auto-shot noise $(S_{LL})$ and cross-shot noise $(S_{LR})$ are essential to characterize the fluctuations of spin current when $B_x$ is present. Moreover, our model calculations show that the sign of the cross-shot noise of spin current is negative for all surface states of AS/QD/AS junctions, while it oscillates between positive and negative values for two surface states of AB/QD/AB junctions as we sweep the gate voltage.
\end{abstract}
%\pacs{72.25.-b, 72.70.+m, 85.35.-p,  85.35.Be }
%Use showkeys class option if keyword

\maketitle

%############################################################
%######################                        ##########################
%######################      Section I         ##########################
%######################                        ##########################
%#############################################################
\section{INTRODUCTION}\label{sec1}
Shot noise describes the fluctuations of current, and It is an inherent characteristic of nano-devices because of the quantization of electron charge. Unlike for thermal noise which describes the equilibrium property of fluctuation in the occupation number of discrete charged particles  $\big<n\big>$, we have to investigate the non-equilibrium (transport) state of the system to observe the shot noise. Generally, the thermal noise is directly dependent on  temperature and gives rise to the occupation number of the states of a system to fluctuate. During the past two decades, the study of shot noise has attracted increasing attention both experimentally and theoretically \cite{a1} since it can give us more detailed information about transport features compared to that of the current.
 \par
 Recently, as a result of the revolution of spinotronic, spin-polarized current particularly pure spin current has received much more attention\cite{a2}. More attention has been paid on the charge current correlation compared with the polarized spin current correlation \cite{a3,a4,a5,a6,a7,a8}. Because of the discreteness of the spin carrier, the information related to spin can be derived from the spin current fluctuations. Shot noise of spin-polarized current has been investigated in several quantum devices containing N-M-N (normal-magnetic-normal)\cite{a9} and F-N-F(Ferromagnet-Normal-Ferromagnet)\cite{a10}. In these devices, shot noise is expected to provide extra information about spin accumulation and spin-dependent scattering process. As It is clear, the charge current correlation between different contacts (cross correlation noise) is definitely negative for a two-contact normal system,\cite{a11}. On the other side, the spin cross correlation noise between different contacts is not necessarily negative for a magnetic connection because of spin flip mechanism. For instance, Ref.[12] showed that the cross correlation at specific Fermi energy can be positive in view of Rashba interaction.
\par
Despite some theoretical and experimental efforts on systems with two metal contacts\cite{a13,a14,a15,a16,a17,a18,a19,a20,a21,a22}, often gold, intriguing physics occurs when one or both contacts are substituted by semiconductors contacts\cite{a23,a24,a25,a26,a27,a28,a29,a30} which includes negative differential resistance and rectification\cite{a31,a32}. Using scattering matrix theory and a generalized Green's function method in this present article, we studied the effect of both spin and charge shot noise on electronic conductance of the scattering QD connected to two semiconductor contacts including silicon and titanium dioxide which has not been studied so far.
\par
The layout of this paper is as follows. In section $(II)$, we summarize the model. Theory and formalism are given in section $(III)$. More details of current fluctuations are presented in Supporting information. In section $(IV)$, we present our numerical results and finally, conclusion is given in section $(V)$.
%############################################################
%######################                        ##########################
%######################      Section II         ##########################
%######################                        ##########################
%#############################################################
\section{Model}\label{sec2}
\begin{figure}
 \includegraphics[width=0.7\columnwidth]{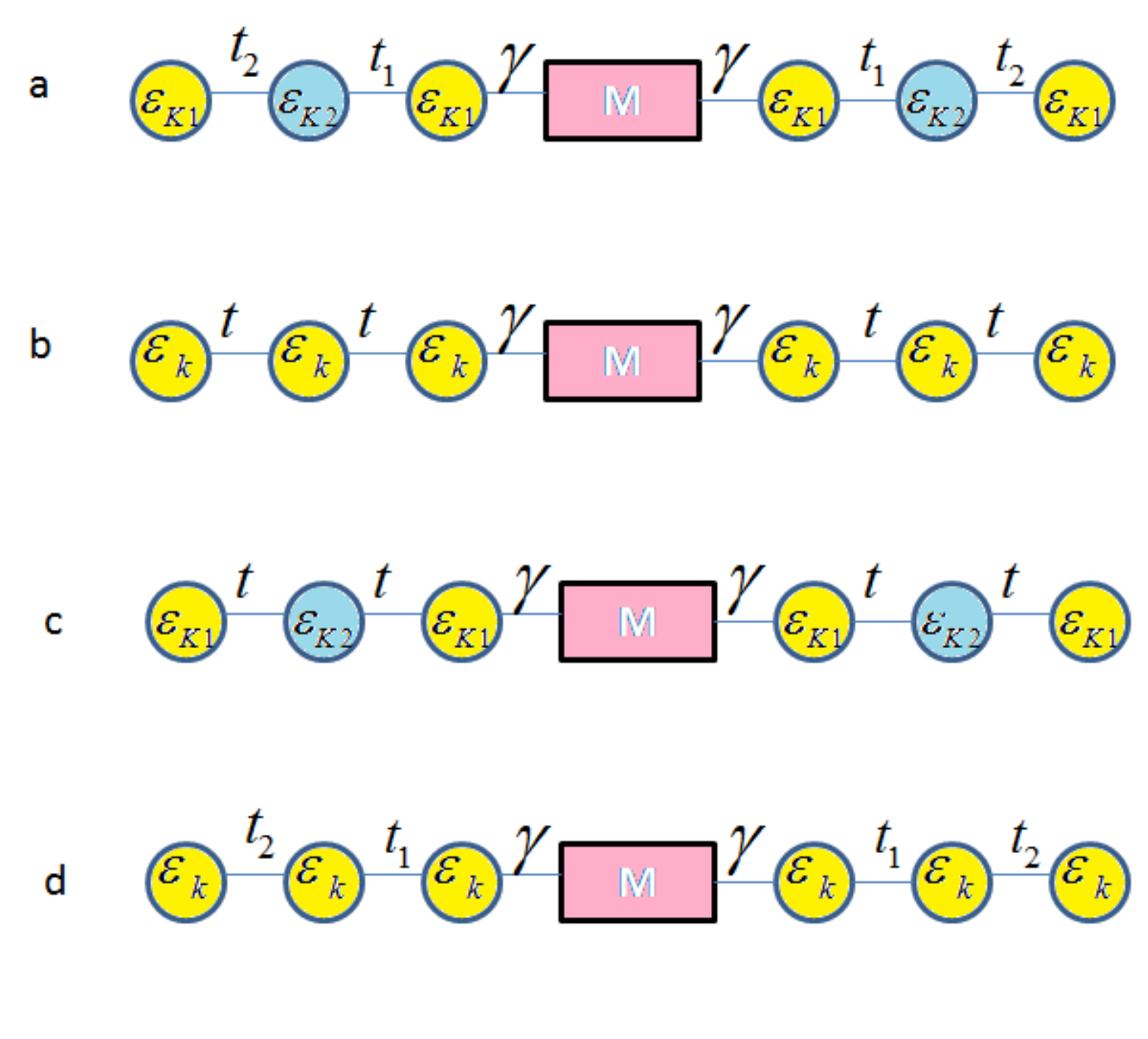}
 \caption{Schematic representation of (a) Koutecky-Davison (KD), (b) Newns-Anderson (NA),
 (c) Alternative site (AS) and (d) Alternative bond (AB) models.}
\label{fig1}
\end{figure}
\par
Using some perturbations of the Newns-Anderson metal model like gold, one can  model semiconductors as titanium-dioxide and silicon \cite{a33,a34,a35,a36,a37,a38} as seen in Fig.1. One such model which alternates both the site energies ${\varepsilon_k}_1$, ${\varepsilon_k}_2$ and inter site couplings $t_1, t_2$ with nearest neighbour in tight-binding picture has been introduced by Koutecky and Davison (KD)\cite{a33}, Fig. 1-(a). Note that, the KD model has three limits. The first one can be achieved in the mixed limit $\varepsilon_k\rightarrow0$, $t_1\rightarrow t_2\equiv t$, producing the Newns-Anderson model (NA) as can be seen in Fig.1-(b) where $\varepsilon_k$ and $t$ are the on-site energy and intersite hopping energies, respectively, in the tight-binding model \cite{a32}. The second type is the limit ${\varepsilon_k}_1$ = $-{\varepsilon_k}_2$, $t_1\rightarrow t_2\equiv t$. This alternating site (AS) model has been used to model the titanium dioxide where the site energies $(\varepsilon_k$ and $-\varepsilon_k)$ refer to the different atoms. The last one is the model alternates bonds (AB) Fig.1-(d). The AB model has been used to model germanium and silicon where the bond disparities $(t_1)$ and $(t_2)$ are related to the orbital hybridization\cite{a32}.
\par
 Cleaving a crystal into two non-interacting parts, leads to symmetries break and surfaces arises. The surfaces show dangling bonds potentially leading to reconstructing surface states with densities localized near the surface \cite{a32,a39,a40}. Three 0, 1 and 2 surface states are made in either AS or AB models. For the AB model, 0, 2 and 1 surface states are characterized as $|t_1| >|t_2|$ in both leads, $|t_1| < |t_2|$ in both leads and $|t_1| > |t_2|(|t_1| < |t_2|) $ in source (drain), respectively. For the AS model, 0, 2 and 1 surface states are defined as ${\varepsilon_k}_1 < 0$ in both leads, $ {\varepsilon_k}_1 > 0$ in both leads and ${\varepsilon_k}_1 < 0$ (${\varepsilon_k}_1 > 0$) in source (drain), respectively.
%############################################################
%######################                        ##########################
%######################      Section III         ##########################
%######################                        ##########################
%#############################################################
\section{Theory formalism}\label{sec2}
The concept of the scattering method is to explain transport properties especially current fluctuations of the system in terms of its scattering properties.
To be clear, we regard a sample connected to electron baths via a number of contacts labeled by an index $\alpha$, and particles which obey Fermi distribution functions $f_\alpha(E)$ = ${[exp[(E-\mu_\alpha)/k_BT_\alpha]}+1]^{-1},\alpha$ = $1,2,3,...$, where $ T_{\alpha}$ and $\mu_{\alpha}$ correspond to temperature and chemical potential of wide leads, respectively. At a given energy $E$ the lead $\alpha$ supports $M_\alpha(E)$ transverse channels. Notice that, each spin degree of freedom is individually investigated. Now, we introduce creating and annihilating operators as ${\hat{a}^\dag}_{\alpha m}({\hat{b}^\dag}_{\alpha m})$ and $\hat{a}_{\alpha m}(\hat{b}_{\alpha m})$ = $\hat{a}_{\alpha m\uparrow}(\hat{b}_{\alpha m\uparrow})$+$\hat{a}_{\alpha m\downarrow}(\hat{b}_{\alpha m\downarrow})$ ($m$ = $1,2...,M_\alpha$), which describe electrons in incoming (outgoing) state of lead $\alpha$ in transverse channel $m$, respectively. $\hat{a}$ and $\hat{b}$ are related through the spin-dependent scattering matrix which is $\hat{b}_{\alpha\sigma} =\sum_{\beta{\sigma}^\prime}s_{\alpha\beta\sigma{\sigma}^\prime}\hat{a}_{\beta{\sigma}^\prime}$, where $\sigma$ and $\sigma^\prime$ are spin variables and $\hat{b}_\alpha$ is a vector of the operator $\hat{b}_{\alpha m}$. Here the indices $\alpha$ and $\beta$ label the contacts. The matrix $\textbf{s}$ is unitary. Using scattering matrix, we define the expression of the charge and spin current operator as follows: \cite{a1}
\begin{eqnarray}
  \hat{I}_{\alpha\mu}(\omega)=&-&\frac{1}{\hbar}\int{dE[\hat{a}^\dag_\alpha(E)\hat{O}_\mu \hat{a}_\alpha(E') }
  -\hat{b}^\dag_\alpha(E)\hat{O}_\mu \hat{b}_\alpha(E')]\nonumber\\
  &-&\frac{1}{\hbar}\int{dE \sum_{\beta\gamma}{\hat{a}^\dag}_\beta(E)A_{\beta\gamma,\mu}(\alpha,E,E')\hat{a}_\gamma(E')}
  \label{e1}
\end{eqnarray}
 Note that, $A_{\beta\gamma,\mu}(\alpha,E,E')$ = $1_\alpha\delta_{\alpha\beta}\delta_{\alpha\gamma}\hat{O}_\mu-{s^\dag}_{\alpha\beta}(E)\hat{O}_\mu s_{\alpha\gamma}(E')$, where $E'$ = $E+\hbar\omega$. Here $\hat{I}_{\alpha \mu}$
with $\mu$ = $1, 2, 3$ defines the spin current operator when
$\hat{O}_\mu$ = $(\hbar/2)\sigma_\mu$ and $\hat{I}_{\alpha4}$ corresponds to the charge current
operator which {$\hat{O}_4$ = $-eI$}, where $I$ is the identity matrix. From Eq.(1), we can derive the average current as $I_{\alpha\nu}$ = ${-1/\hbar}\int {dE\sum_{\beta}Tr[A_{\beta\beta,\mu}(\alpha)]f_\beta(E)}$.
\par
It is straightforward to show that the spectral density of current fluctuations $S_{\alpha\beta}(\omega)$ in terms of its Fourier-transformed current operator is $2\pi\delta(\omega+\omega')S_{\alpha\beta}(\omega)\equiv\big<\Delta\hat{I}_\alpha(\omega)\Delta\hat{I}_\beta(\omega')+\Delta \hat{I}_\beta(\omega')\Delta
  \hat{I}_\alpha(\omega)\rangle$, where $\Delta \hat{I}_{\alpha}(\omega)\equiv \hat{I}_{\alpha}(\omega)-\big< \hat{I}_{\alpha}(\omega)\big>$.
%%%%%%%%%%%%%%%%%%%%%%%%%%%%%%%%%%%%%%%%%%%%%%%%%%%55555checked up
Hence, current fluctuations is described as,
\begin{eqnarray}
\big<\Delta\hat{I}_{\alpha\mu}(\omega)\Delta\hat{I}_{\beta\nu}(\omega')\big>=\frac{1}{\hbar^2}\sum_{\delta\gamma}\int{dETr[A_{\delta\gamma,\mu}(\alpha,E,E')}\nonumber\\
\times A_{\delta\gamma,\upsilon}(\beta,E',E)]f_{\gamma}(E)[1-f_{\delta}(E')]\delta(\hbar\omega+\hbar\omega')
\label{e9}
\end{eqnarray}
 For further discussion, please see Eqs.(1-5) of supporting information.
 $\big<\Delta\hat{I}_{\alpha\mu}\Delta\hat{I}_{\beta\nu}\big>_{\omega}\equiv\Delta\upsilon S_{\alpha\beta}(\omega)$, where $\Delta\upsilon$ is a frequency interval\cite{a41}. In the low-frequency limit, shot noise is given as,
 \begin{eqnarray}
S_{\alpha\beta,\mu\nu}&=&\frac{1}{\pi\hbar}\sum_{\gamma\delta}\int{dEf_\gamma(1-f_\delta)}\nonumber\\
&\times&Tr[A_{\gamma\delta,\mu}(\alpha,E,E)A_{\delta\gamma,\nu}(\beta,E,E)]
\label{e10}
\end{eqnarray}
\par
We can rewrite the above equation in terms of the scattering matrix and Fermi distribution (please see Eqs.(6-8) of the supporting information). It is straightforward to show that $\sum_\alpha I_{\alpha 4}$ = $0$ because $\sum_\alpha A_{\beta\gamma,4}(\alpha,E,E)$ = $0$. Therefore, both the charge current as well as the noise of charge current are conserved quantities. Even though $\sum_\alpha A_{\beta\gamma,i}(\alpha,E,E)\neq0$, we still have $\sum_\alpha Tr[A_{\beta\gamma,i}(\alpha,E,E)]$ = $0$ for $\hat{O}_\mu$ = $(\hbar/2)\sigma \mu$. Hence, the total spin current is a conserved quantity again $\sum_\alpha {{I_{\alpha_3}}}$ = $0$. Notice that, the spin shot noise is not a conserved quantity due to the relation which is $\sum_\alpha A_{\beta\gamma,i}(\alpha,E,E)\neq0$. So the auto correlation and the cross correlation for spin current we will describe below, are two apart quantities and are independent of each other.
\par
Generally, there are two different kinds of fluctuations in detail. The first one is the auto-correlation that is fluctuations between currents in the same contact; the second one is cross-correlation that is fluctuations between distinct contacts. Below we address these different fluctuations.
\par
\textbf{\emph{Auto correlation}:}Substituting $\beta$ for $\alpha$ in Eq.(3), we find the general expression for the auto correlation at prob $\alpha$;
 \begin{eqnarray}
\emph{S}_{\alpha\alpha,\mu\nu}=\frac{2}{h}\int{dE \Big(f_\alpha (1-f_\alpha)\Big[Tr(\hat{O}_\mu 1_\alpha \hat{O}_\nu)-Tr(\hat{O}_\mu s^\dag_{\alpha\alpha}}\nonumber\\
 \times \hat{O}_\nu s_{\alpha\alpha})-Tr(s^\dag_{\alpha\alpha} \hat{O}_\mu s_{\alpha\alpha}\hat{O}\nu)\Big]+\sum_{\gamma}f_{\gamma}Tr(\hat{O}_\nu s_{\alpha\gamma}\nonumber\\
\times s^\dag_{\alpha\gamma}\hat{O}_\mu)-\sum_{\gamma\delta}f_\gamma f_\delta Tr(s^\dag_{\alpha\gamma}\hat{O}_\mu s_{\alpha\delta}s^\dag_{\alpha\delta}\hat{O}_\nu s_{\alpha\gamma})\Big)
\label{e12}
\end{eqnarray}
\textbf{\emph{Cross correlation}:}
For $\alpha\neq\beta$, the cross correlation of the current fluctuations at two terminals is described as;
\begin{eqnarray}
S_{\alpha\beta,\mu\nu}&=&\frac{2}{h}\int{dE \Big(f_\alpha (1-f_\alpha)Tr(\hat{O}_\mu s^\dag_{\beta\alpha}\hat{O}_\nu s_{\beta\alpha})}\nonumber\\
&+&f_\beta (1-f_\beta)Tr(s^\dag_{\alpha\beta} \hat{O}_\mu s_{\alpha\beta}\hat{O}\nu)\nonumber\\
&-&\sum_{\gamma\delta}[f_\gamma f_\delta Tr(s^\dag_{\alpha\gamma}\hat{O}_\mu s_{\alpha\delta}s^\dag_{\beta\delta}\hat{O}_\nu s_{\beta\gamma})]\Big)
\label{e13}
\end{eqnarray}
 The shot noise at zero temperature is calculated by subtracting the equilibrium-like auto correlation and equilibrium-like cross correlation from the general expression of auto and cross correlation, respectively. In the derivation of shot noise, we essentially follow supporting information (section 3, 4 and 5). For a two contact device, It is as follows: \cite{a12},
\begin{eqnarray}
\emph{S}_{LL,\mu\nu}=\frac{2}{h}\int{dE\Big\{(f_L-f_R)^2Tr[s_{LL}s_{LL}^\dag \hat{O}_\mu s_{LR}s_{LR}^\dag \hat{O}_\nu]}\nonumber\\
+f_L(1-f_L)[\hat{O}_\mu s_{LL} s_{LL}^{\dag}\hat{O}_\nu-\hat{O}_\mu s_{LL}^{\dag} \hat{O}_\nu s_{LL}]\nonumber\\
+f_L(1-f_L)[\hat{O}_\nu s_{LL} s_{LL}^{\dag}\hat{O}_\mu-\hat{O}_\nu s_{LL}^{\dag} \hat{O}_\mu s_{LL}]\nonumber\\
+[(f_L(1-f_L)+f_R(1-f_R))Tr[\hat{O}_\mu s_{LR}s_{LR}^\dag \hat{O}_\nu ]\Big\}
\label{e16}
\end{eqnarray}
From Eq.(3), the cross-shot noise for a two-terminal system is described as,
\begin{equation}\label{e19}
  \emph{S}_{LR,\mu\nu}=-\frac{2}{h}\int{dE\{(f_L-f_R)^2Tr[s_{LL}^\dag \hat{O}_\mu s_{LL}s_{RL}^\dag \hat{O}_\nu s_{RL}]}\}
\end{equation}
 The first term in  Eq.(6) is the shot noise and the other three terms are the thermal (equilibrium) fluctuations as a result of fluctuations in the occupation numbers of the incident channels, disappearing at zero temperature.
\par
Because the auto correlation is definitely positive, the cross shot noise for charge current must be negative as a result of conservation law which is described as $S_{LL}$ = $-S_{LR}$. Nevertheless, the cross-shot noise of spin current may be supposed to be a positive value, the spin current fluctuations is not a conserved quantity.
\par
Using the Fisher-lee relation\cite{a42} which is defined as $s_{\alpha\beta}$ = $-\delta _{\alpha \beta}+i \Gamma_\alpha ^{1/2} G^r \Gamma_\beta ^{1/2}$, we obtain the different noise expressions in terms of the Green’s function.
\par
In the coherent tunnelling regime, the QD-surface interactions are indirectly described by the effect they have on the QD, which is formally performed by a self-energy as $\Sigma(E)$ = $\Lambda(E)-\frac{i}{2}\Gamma(E)$. $\Lambda(E)$ is the real part of self-energy corresponding to the shift of a energy levels. In the wide band limit this energy shifting is trivial for metal and can be ignored\cite{a32}, while this is  non-trivial and remarkable for semiconductors. Imaginary part $\Gamma(E)$ of the self-energy defines the broadening of the energy levels originated by electrode. Here, we consider a QD in the presence of two external magnetic fields $B_x$ and $B_z$. The supposed QD has only a single state $|s \rangle$ of energy $\epsilon$. On the other hand, the “atomic” levels of a particular contact is denoted by $|a_j\rangle$ and the Bloch states of the same contact by $|k\rangle$ with corresponding energies $\epsilon_k$ \cite{a32}. The QD state only pairs to the terminal atomic level, $|a_1\rangle$, in the tight-binding framework in which each contact site has solely a single level, in other words, $\langle s|H|a_j\rangle$ = $\gamma\delta_{j,1}$, where $H$ is the total Hamiltonian of the system in which the QD connected by two infinite electrodes $H$ = $H_{L(R)}+H_{T}+H_{QD}$. The first term $H_{L(R)}$ defines the Hamiltonian of the left (right) contact which we consider as the following Hamiltonian known as $KD$ model,
\begin{eqnarray}
\emph{H}_{KD}&=&\varepsilon_k\sum_{j=1}^N\big(a_{(2j-1)}^\dag a_{(2j-1)}-a_{2j}^\dag a_{2j}\big)
+\Big[t_1\sum_{j=1}^{N}\big(a_{(2j-1)}^\dag \nonumber\\
&\times&a_{2j}\big)
+t_2\sum_{j=1}^{N-1}\big(a_{(2j+1)}^\dag a_{2j}\big)+h.c\Big]
\label{e2}
\end{eqnarray}
where $a^\dag_j$ creates an electron at site $j$ in this contact. As is stated before there are three important limits of the $KD$ model as the $NA$, $AS$ and  $AB$ models which their Hamiltonian can be expressed as
  $\emph{H}_{NA}$ = $t\sum_{j=1}^ N a_{j} ^ \dag a_{j}+h.c$, $\emph{H}_{AS}$ = $\varepsilon_k\sum_{j=1}^N\big(a_{2j-1}^ \dag a_{2j-1}-a_{2j}^\dag a_{2j}\big)
+t\sum_{j=1} ^{2N-1}a_{j} ^\dag a_{j}+h.c$ and $\emph{H}_{AB}$ = $t_1\sum_{j=1} ^N a_{2j-1}^ \dag a_{2j}-t_2\sum_{j=1}^{N-1} a_{2j+1} ^\dag a_{2j}
+h.c$, respectively \cite{a31}.
The second term $H_{T}$ in the whole Hamiltonian of system is the Hamiltonian which describes the coupling between the isolated QD and contacts
with the coupling parameter $\gamma_{j\sigma}$, which is described as $\emph{H}_T$ = $\sum_{j\sigma}\gamma_{j\sigma}\big(a_j^\dag C_{\sigma}+h.c\big)$,
where h.c. is referred to the complex conjugate. The last term in the total Hamiltonian expression is the isolated $QD$ Hamiltonian under two transverse $(B_x)$ and longitudinal $(B_z)$ external magnetic fields $H_{QD}$ = $\sum_{\sigma}\epsilon C^\dag_{\sigma}C_{\sigma}+\sum_{\sigma}C^\dag_{\sigma}[B_x\sigma_x+B_z\sigma_z ]C_{\sigma}$, where $C^\dag_{\sigma}(C_{\sigma})$ denotes the creation (annihilation) operator of an electron at site $i$ and $\sigma_x$($\sigma_z$) are pauli matrices in different directions.
\par
The spectral density which is the result of adsorption, for isolated resonances is given by $\Gamma_{L(R)}(E)$ = $2\pi\sum_K|V_k|^2\delta(E-\varepsilon_k)$, where $V_k\equiv<s|H|k>$ = $\gamma\langle a_1|k\rangle$. Having the eigenvectors and the eigenvalues \cite{a32}, the spectral density can be obtained from the above equation as
\begin{eqnarray}
\Gamma_{KD}(E)&=&\frac{\gamma^2}{t_2^2}\Big(\big[{\varepsilon_k}^2+(t_1+t_2)^2-E^2\big]\big[E^2-{\varepsilon_k}^2-(t_1\nonumber\\
&-&t_2)^2\big]\Big[\frac{1}{(E-\varepsilon_k)^2}\Big]\Big)^{1/2}
\label{e31}
\end{eqnarray}
[For more discussion please see section (6) of supporting information].
%\end{widetext}
For $[{\varepsilon_k}^2+(t_1-t_2)]^2]^{1/2}\leq|E|\leq[{\varepsilon_k}^2+(t_1+t_2)]^2]^{1/2}$, where $\varepsilon_k$ is the contact state energy. Consider that, $\Gamma_{L(R)}(E)$ can be related to $\Lambda_{L(R)}(E)$ through the Hilbert transform, when $\Sigma_{L(R)}(E)$ has no singularities
on the real energy axis.
%%%%%%%%%%%%%%%%%%%%%%%%%%
\begin{eqnarray}
\frac{\Lambda_{KD}(E)}{\gamma^2}&=&{E^2-\varepsilon_k^2-t^2_1+t^2_2+\Theta_{KD}(E)\Big(\big[E^2-{\varepsilon_k}^2-(t_1}\nonumber\\
&-&t_2)^2\big] \big[E^2-{\varepsilon_k}^2-(t_1+t_2)^2\big]
\Big[{{\frac{1}{2t_2^2(E-\varepsilon_k)}\Big]}\Big)^{1/2}}\nonumber\\
 \Theta_{KD}(E)&=&\Theta({\varepsilon_k}^2+(t_1-t_2)^2-E^2)-\Theta(E^2-{\varepsilon_k}^2\nonumber\\
 &-&(t_1+t_2)^2),
\label{e32}
\end{eqnarray}
%%%%%%%%%%%%%%%%%%%%%%%%%%%%%%%%%
Consider that $\Lambda_{KD},\Gamma_{KD}\rightarrow\Lambda_{NA},\Gamma_{NA}$, by applying $\varepsilon_k\rightarrow0$, $t_1\rightarrow t_2\equiv \beta$. Also taking the limits $\alpha\rightarrow0$ and $t_1\rightarrow t_2\equiv t$ gives the relations for the $AB$ and $AS$ models, respectively. The more detailed information is  presented in the supporting information document.
\par
The transmission spectrum $\mathcal{T}$ is defined as $\mathcal{T}$ = $Tr[{\textbf{T}}]$, where ${\textbf{T}}$ = $s_{LR}^\dag s_{LR}$. The matrix of
the retarded green function is defined as $\textbf{G}^\textbf{R}$ = $1/{[E\textbf{I}-\textbf{H}-\Sigma_L-\Sigma_R]}$, where $\textbf{I}$ stands for the identity matrix.\\
\par
There are three kinds of different fluctuations which are remarkable. The first one is the current-current correlation in the same contact or distinct contacts. The second kind is the current-spin correlation in the same or different contacts with various spin directions. The last type is the spin-spin correlation including fluctuations for spin along different directions and between different contacts. In this paper, we just take care the last one and we only consider the correlation with the same spin direction $\sigma_3$. (consistently $\mu$ = $1, 2, 3$)
\par
 The spin current is described as follows:
\begin{equation}\label{e42}
  I_{L3}=\frac{1}{2}\int{\frac{dE}{2\pi}Tr[{\sigma_3 \textbf{T} ](f_L-f_R)}}
\end{equation}
Note that, for the QD with one level, the spin current is, in general, non-zero. The auto-shot noise and cross-shot noise of spin current for a two-terminal system were discussed in section $II$.
%############################################################
%######################                        ##########################
%######################      Section IV         ##########################
%######################                        ##########################
%#############################################################
\section{Numerical result}
In this section, based on the theory formalism described in section $(II)$, we present numerical calculations of charge-spin current and cross-auto shot noise versus gate voltage through an isolated QD with a resonant level coupled with two metal/semiconductor electrodes, where two $B_x$ and $B_z$ magnetic fields are applied to QD. As a essential basis for comparison between semiconductors and metal electrodes, we first consider a gold/QD/gold junction within the NA model. The values of NA, AB and AS models parameters are listed in Table-I. As a reference energy, the Fermi energy of electrodes is fixed at $E_F=0$. The temperature is also fixed at $T = 4K$.
%############################################################
%######################                        ##########################
%######################      Section IV-a         ##########################
%######################                        ##########################
%#############################################################
\subsection{Metal electrode}
\begin{figure}
 \includegraphics[width=1.\columnwidth]{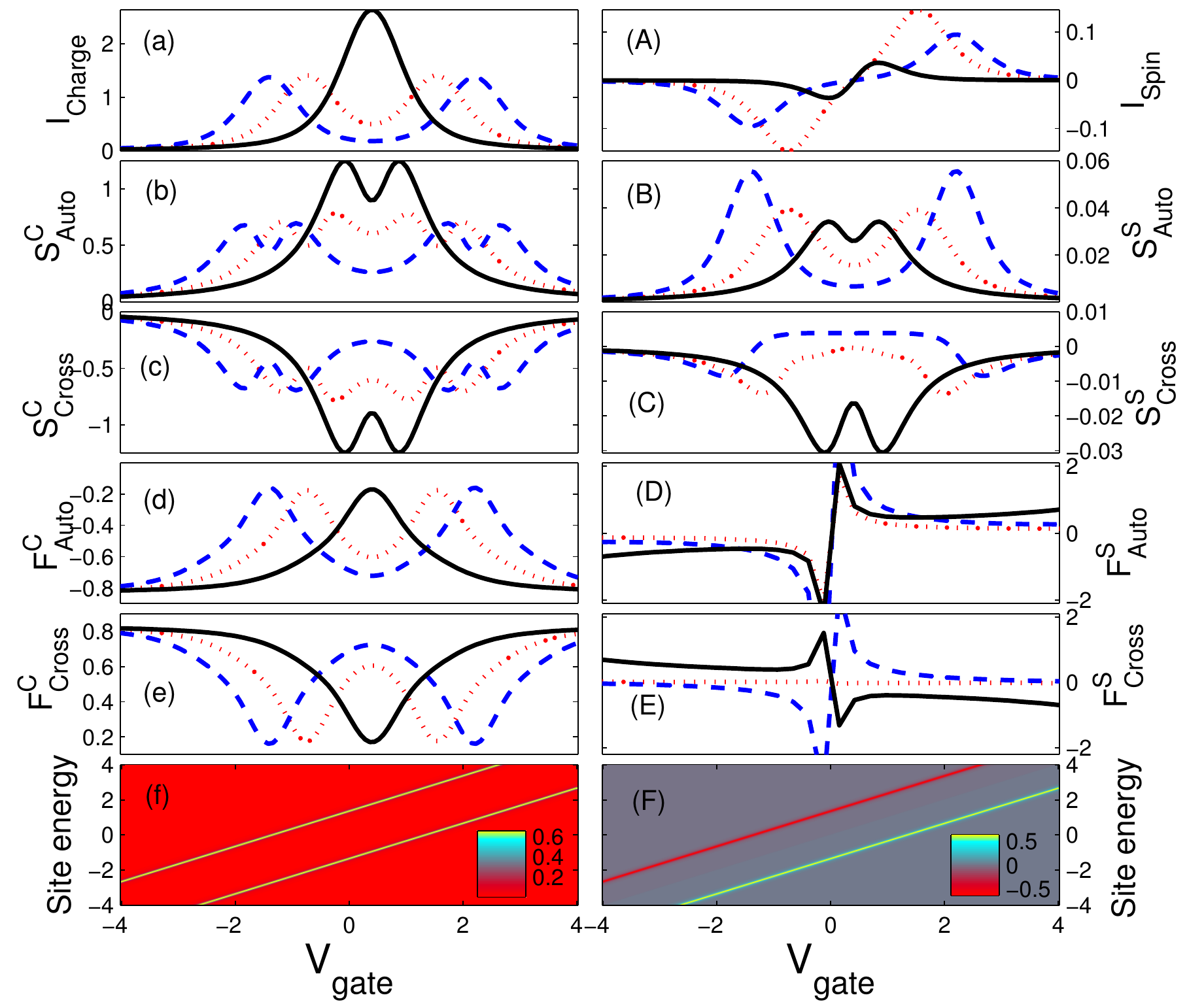}
\caption{Panels (a), (b), (c), (d) and (e) provide the information about the charge current, charge auto-shot noise, charge cross-shot noise, charge auto-fano and charge cross-fano versus gate voltage of the metallic junction, for different parameters: (1)  $B_z=0.1$, $B_x=0.1$ (dot red line)  (2)  $B_z=0.1$, $B_x=0.2$ ( dashed blue line)  (3) $B_z=0.01$, $B_x=0.01$ ( solid black line), respectively (here we fixed site energy as $\epsilon=0.05$). Panels labelled with capital letters have the same meaning for spin case.  Panels (f) and (F) are a density plot correspond to the charge  and spin current through the junction for various molecular site energies and applied gate voltages.  We fixed magnetic fields as $B_z=1$, $B_x=0.9$}
\label{fig2}
\end{figure}

 The left(right) panels of Fig.~\ref{fig2} show the  charge(spin) current, auto-shot noise, cross-shot noise, auto-fano and  cross-fano versus gate voltage of the gold/QD/gold junction.  It is obvious that, there is only one energy level, $\epsilon$, in the QD: after diagonalizing the Hamiltonian, this level is divided into two levels, $\epsilon\pm\sqrt{{B_{x}}^2+{B_{z}}^2}$, in which resonant peaks rise. As understood from Fig.~\ref{fig2}-(a), the number of peaks differs for various parameters, which can be described in the language of resonant states. When $\sqrt{{B_{x}}^2+{B_{z}}^2}\ll\epsilon$, because of overlapping of the two resonant levels, only one sharp peak occurs (solid black line). If $\sqrt{{B_{x}}^2+{B_{z}}^2}>\epsilon$, two peaks appears (dotted red line and dashed blue line). Fig.~\ref{fig2}-(A) shows that the spin current changes its sign as we sweep the gate voltage.
%%%%%%%%%%%%%%%%%%%%%%%%%%%%%%%%%%%%%%%%
%%%%%%%%%%%%%%%%%%%%%%%%%%%%%%%%%%%%%%%
\begin{table}[b]
\caption{\label{table1} Model Parameters for Au, Si, and TiO$_2$}
\begin{ruledtabular}
\begin{tabular}{ccdddd}
material&model&
\multicolumn{1}{c}{\textrm{$|\varepsilon_{k}|(eV)$}}&
\multicolumn{1}{c}{\textrm{$t_1(eV)$}}&
\multicolumn{1}{c}{\textrm{$t_2(eV)$}}&
\multicolumn{1}{c}{\textrm{$\gamma(eV)$}}\\
%\mbox{Three}&\mbox{Four}&\mbox{Five}\\
\hline
Au\cite{a11}&NA& &-8.95& & -0.45\\
Si&AB&  & -1.60 & -2.185 & -1.0 \\
TiO$_2$ &AS & 1.6 & -2 &  & -1.0 \\
\end{tabular}
\end{ruledtabular}
\end{table}
%%%%%%%%%%%%%%%%%%%%%%%%%%%%%%%%%%%%%%%%
%%%%%%%%%%%%%%%%%%%%%%%%%%%%%%%%%%%%%%%
% ================================\\
% ================================\\
% ================================\\
 %================================
 Fig.~\ref{fig2}-(b) and (c) show charge auto-shot noise and cross-shot noise. As it can be seen, the general trend is the same as charge current but now each peak comes in pairs  which signals each spin-up and spin-down contribution to the total charge current noise. They do not show any sign changing with choosing different parameters. Actually, the charge correlation between different (same) probes is negative (positive) for fermions which obvious here. There is also a very nice mirror symmetry, respect to zero line which is consistent with this relation $\sum_{\alpha}^e S_{\alpha\beta}$ = $\sum_{\beta}^e S_{\alpha\beta}$ = $0$. Fig.~\ref{fig2}-(B) and Fig.~\ref{fig2}-(C) show spin auto-shot noise and cross-shot noise. We see  for some parameters spin cross correlation oscillates between negative and positive values as we scan the gate voltage. Because of spin-flip process, either spin-up and spin-down electrons contribute to spin current. The cross-shot noise between spin-up (spin-down) electrons is found to be negative definite, but it is positive definite between electrons with different orientation. Indeed, one can write the total charge current noise as $S_{\alpha\beta}^{charge}=S_{\alpha\beta}^{\uparrow\uparrow}+S_{\alpha\beta}^{\downarrow\downarrow}+S_{\alpha\beta}^{\uparrow\downarrow}+S_{\alpha\beta}^{\uparrow\downarrow}$ and the total spin current noise as $S_{\alpha\beta}^{spin}=S_{\alpha\beta}^{\uparrow\uparrow}+S_{\alpha\beta}^{\downarrow\downarrow}-S_{\alpha\beta}^{\uparrow\downarrow}-S_{\alpha\beta}^{\uparrow\downarrow}$. It is clear from the relations that in the absence of spin-flip mechanism we have $S_{\alpha\beta}^{charge}=S_{\alpha\beta}^{spin}$. While, in the presence of spin-flip mechanism  $S_{\alpha\beta}^{charge}\neq S_{\alpha\beta}^{spin}$. The sign of the correlation $S_{\alpha\beta}^{\uparrow\downarrow}$ (bouncing and antibouncing) is  receiving special interest. The competition between these two contributions $S_{\alpha\beta}^{\uparrow\uparrow}$ and $S_{\alpha\beta}^{\uparrow\downarrow}$ gives rise to either a positive or a negative in the spin current noise spectrum.  In spite of more complicated line shape, the spin cross correlation shows degradation at resonance transmission, hence it can be useful in detecting open channel of spin current. This analogous in detecting open channel of charge current which big suppression happens of the charge cross correlation  at resonance transmission.
 % ================================\\
 %================================\\
 %================================\\
\par
To get better insights, we address the Fano factor ($F$) to characterize the deviation of shot noise compared to the Poisson value $S_{poisson}=-2eF I$. Indeed, the zero-frequency shot noise of the charge current in a two terminal non-interaction conductors reaches the maximum value $F=1$ (the Poissonian limit) which the mean occupation of a state is so small, hence the Pauli principle is trivial. On the other hand, there are two more limits known as sub-Poissonian ($F<1$) and super-Poissonian ($F>1$) cases. Here, we examined two Fano factors corresponding to the two cross and auto charge current noises as $F_{Cross(Auto)}^{C}=-\frac{S_{Cross(Auto)}}{2eI_{C}}$. We also adopt this definition for spin current noises $F_{Cross(Auto)}^{C}=\frac{S_{Cross(Auto)}}{2I_{S}}$ which reveals the transported spin unit.  It is already expected for the charge current noise the two definitions of the Fano factor give the same information which it is clear from Fig.~\ref{fig2}-(d) and (e). This is because that the two cross and auto shot noise differs up to a sign, so one can see the two defined Fano factors show exactly the same sign changing. For different parameter cases we haven't reached to the Poissonian limit and a big suppression happens at resonance points which shows qualitatively  an open channel. In the spin case, we expect to find different Fano factor behaviors for two cross and auto cases. Fig.\ref{fig2}-(D) and (E) show auto and cross Fano factors. It is worth to mention that near the point which the spin current changes its sign (zero spin current), one can see a sharp enhancement of the Fano factor up to $F=2$ fro some parameters.  Actually,  it resembles  to the superconductor with $F=2$ which the Cooper pairing  causes an attractive Coulomb interaction between the electrons. But a swift reduction to the sub-Poissonian limit happens with a little deviate from zero point spin current.
\par
 Fig.~\ref{fig2}-(f)  shows the formation of two resonances in each $\epsilon$, as shown by the ridges of high charge current $(HCR)$, while Fig.~\ref{fig2}-(F) shows the formation of a resonance peak and an anti-resonance peak in each $\epsilon$, as indicated by the ridge of high positive spin current $(HPSR)$ and high negative spin current $(HNSR)$, respectively.
%The inset of Fig.(2-e) shows the charge current when we ignore the molecular shift energy $\Lambda_{NA}(E)$. We found out, after %neglecting the $\Lambda_{NA}(E)$ (see inset) the $HCR$ remains linear and are coincident with the $HCR$ in Fig.(2-e). Therefore, the %weak molecular-level shifting can be omitted for junctions with two metal electrodes.
%%%%%%%%%%%%%%%%%%%%%%%%%%%%%%%%%%%%%%%%
%%%%%%%%%%%%%%%%%%%%%%%%%%%%%%%%%%%%%%%%
%############################################################
%######################                        ##########################
%######################      Section IV-b        ##########################
%######################                        ##########################
%#############################################################
\subsection{Semiconductor electrode}
\begin{figure}
 \includegraphics[width=1\columnwidth]{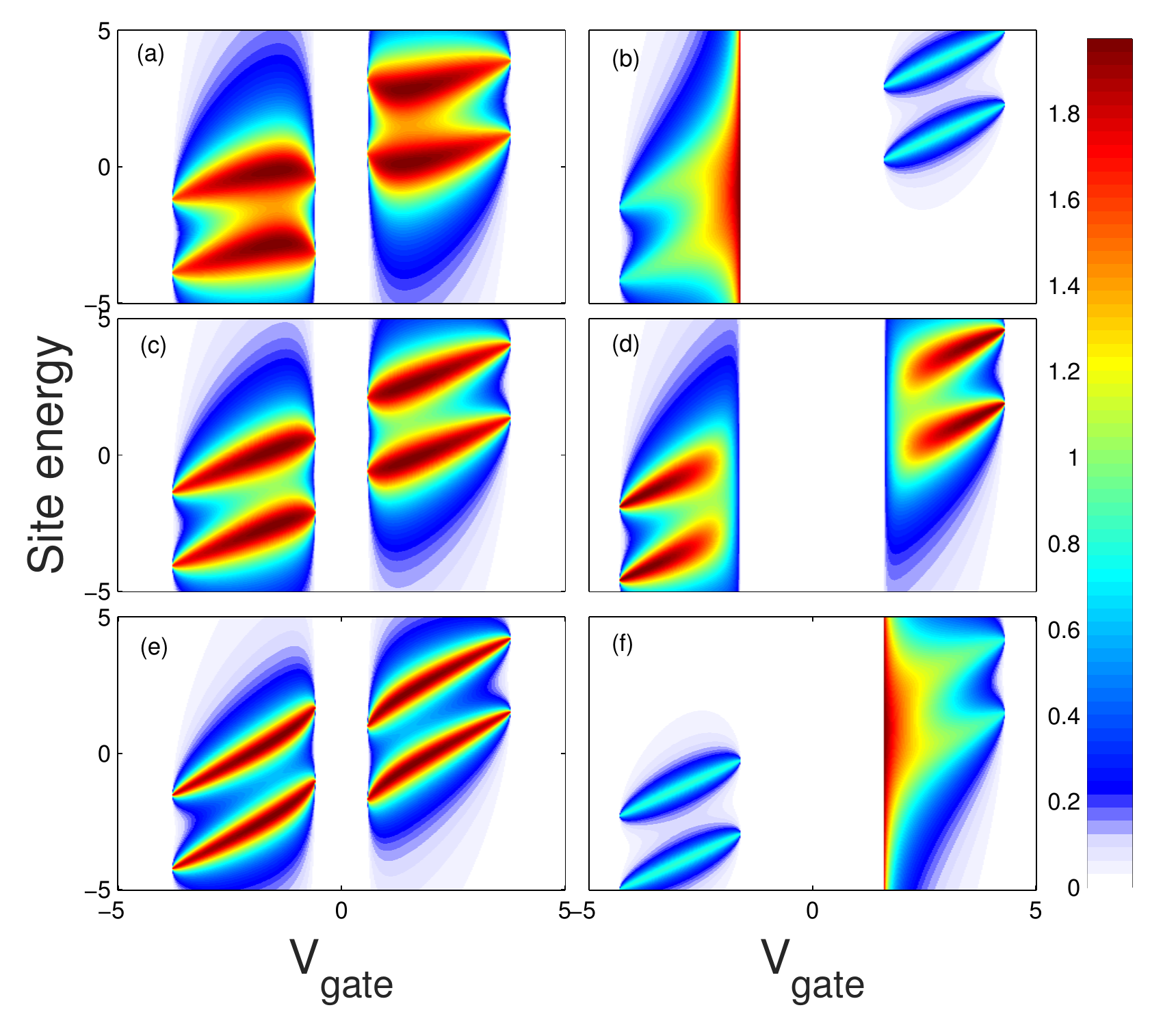}
\caption{The left and the right columns provide information about the charge current  through  AB and AS semiconductor junctions, respectively. Rows from top to bottom are addressing to the 0, 1, and 2 surface states.  We fixed magnetic fields as $B_z=1$, $B_x=0.9$}
\label{fig3a}
\end{figure}

\begin{figure}
 \includegraphics[width=1\columnwidth]{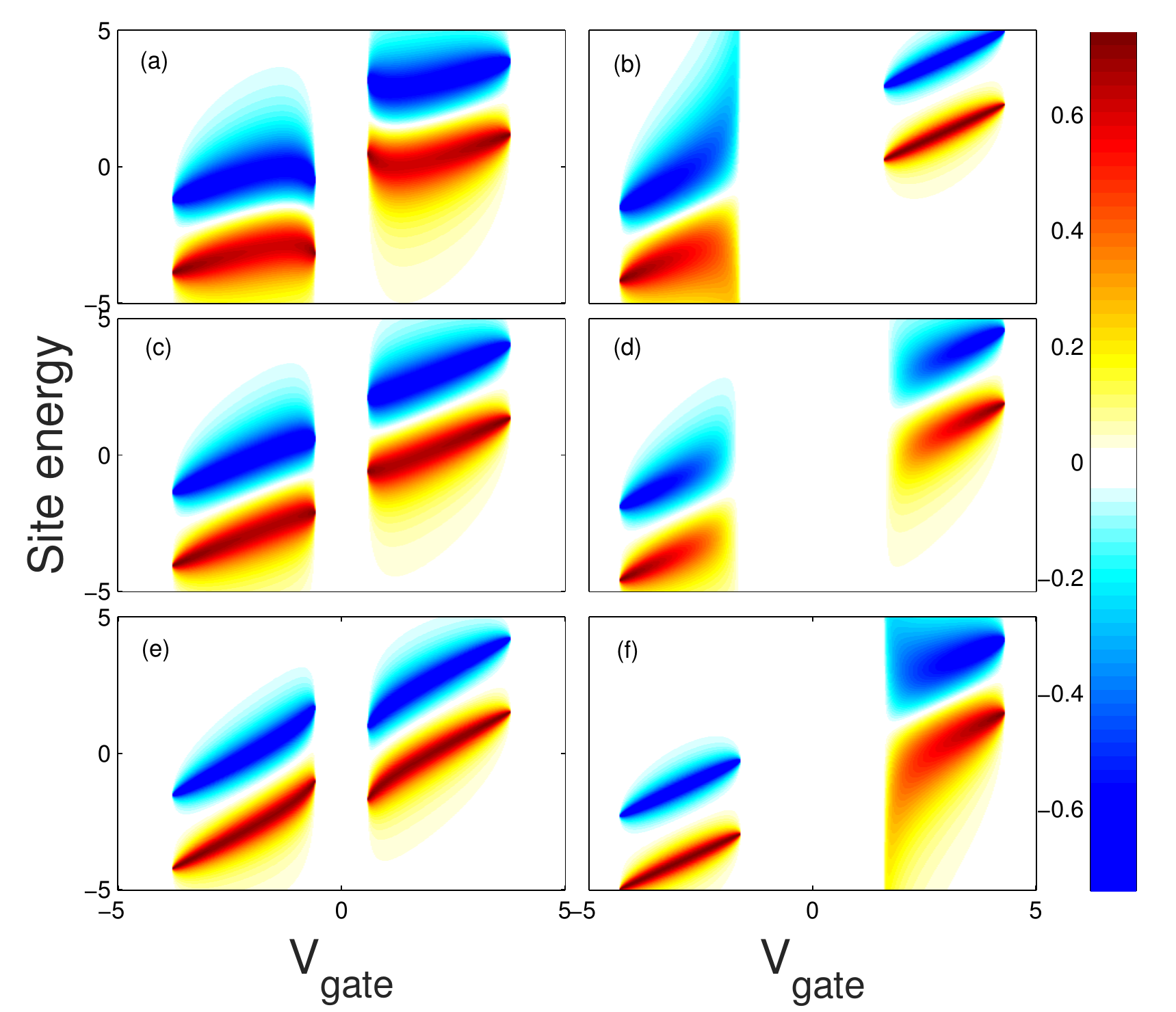}
\caption{The left and the right columns provide information about the spin current  through  AB and AS semiconductor junctions, respectively. Rows from top to buttom are addressing to the 0, 1, and 2 surface states.  We fixed magnetic fields as $B_z=1$, $B_x=0.9$}
\label{fig3b}
\end{figure}

Having computed the different factors versus  gate voltage associated with the NA model, we now proceed to investigate spin-charge current and auto-cross correlation versus gate voltage in terms of systems based on semiconductor electrodes within the AB and AS model. One parameter issue is: how do semiconducting electrodes change the spin-charge current and auto-cross correlation profiles compared to metal electrodes? In the limit of semiconductor/QD/semiconductor junctions, the resonant peaks get substantial widths compared to the metal/QD/ metal connections. The contribution to the widening of the resonant peaks in this junctions limit arises from the molecular shifting energy. This feature is obviously observed by comparing the results plotted in Fig.~\ref{fig2}-(f) and Fig.~\ref{fig2}-(F) with Fig.~\ref{fig3a} and Fig.~\ref{fig3b}. On the other hand, the semiconductor band gap is the most important features of the Fig.~\ref{fig3a} and Fig.~\ref{fig3b}, as shown by the limit of zero spin and charge current in each plot. In these sectors, the absence of states in the left contacts prevents from injecting electrons into the junction; likewise, there are no states for them to occupy once transmitted to the right contact.
 %================================\\
%%%%%%%%%%%%%%%%%%%%%%%%%%%%%%%%%%%%%%%%
%%%%%%%%%%%%%%%%%%%%%%%%%%%%%%%%%%%%%%%%
%%%%%%%%%%%%%%%%%%%%%%%%%%%%%%%%%%%%%%%%
%%%%%%%%%%%%%%%%%%%%%%%%%%%%%%%%%%%%%%%%
%%%%%%%%%%%%%%%%%%%%%%%%%%%%%%%%%
%%%%%%%%%%%%%%%%%%%%%%%%%%%%%%%%%%%%%%%%
\par
 % ================================\\
 %================================\\
 %================================\\
The Fig.~\ref{fig3a} and Fig.~\ref{fig3b} show charge and spin currents  as a function of  gate voltage $V_g$ and on site energy $\epsilon$ for semiconductor junctions, respectively. In the both  Fig.~\ref{fig3a} and Fig.~\ref{fig3b}, left and right column corresponds to the AB/QD/AB and the AS/QD/AS  junctions, respectively.  Rows from top to bottom are addressing to the 0, 1, and 2 surface states. Transverse and longitudinal magnetic fields are set at $B_x$=$0.9$ and $B_z$=$1$.
\par
As it is clear from these figures, the bonding configuration has a noticeable effect on the charge and spin current. The principle effect of bond configuration on the charge is that, the broadening of the HCRs in both bands reduces with more surface state, which is described by the scaling of $\Sigma_{AB}(E)$ with ${t_2}^{-2}$.
%The inset of Fig.(3-a) shows the charge current through its respective connection when the shifts of the molecular energy level, $%\Lambda_{AB}(E)$, is ignored.
$\Lambda_{AB}(E)$, which is significant for semiconductor electrodes, is responsible for both the contorted shape of the HCR as well as its movement away from the diagonal. The electron transport is affected by energy-level shifts and is also sensitive to the presence of the bond configurations. Generally speaking, HCRs are shifted, broadened and bent depending on the presence of bonding configuration (see Fig.~\ref{fig3a}-(a), (c) and (e)). We divide the high spin current into two regions as so called $HNSRs$ and $HPSRs$. The careful inspection of Fig.~\ref{fig3b}-(a), (c) and (e) reveals that the shifting, bending and broadening of $LNSRs$ and $HPSRs$ in both bands are similar to $HCRs$ in  Fig.~\ref{fig3a}-(a), (c) and (e), respectively. The band gap, which is shown by the limit of zero charge and spin current as well as the bonding configurations in these junctions are noticeable.\\
%%%%%%%%%%%%%%%%%%%%%%%%%%%%%%%%%%%%%%%%
\begin{figure}
 \includegraphics[width=1\columnwidth]{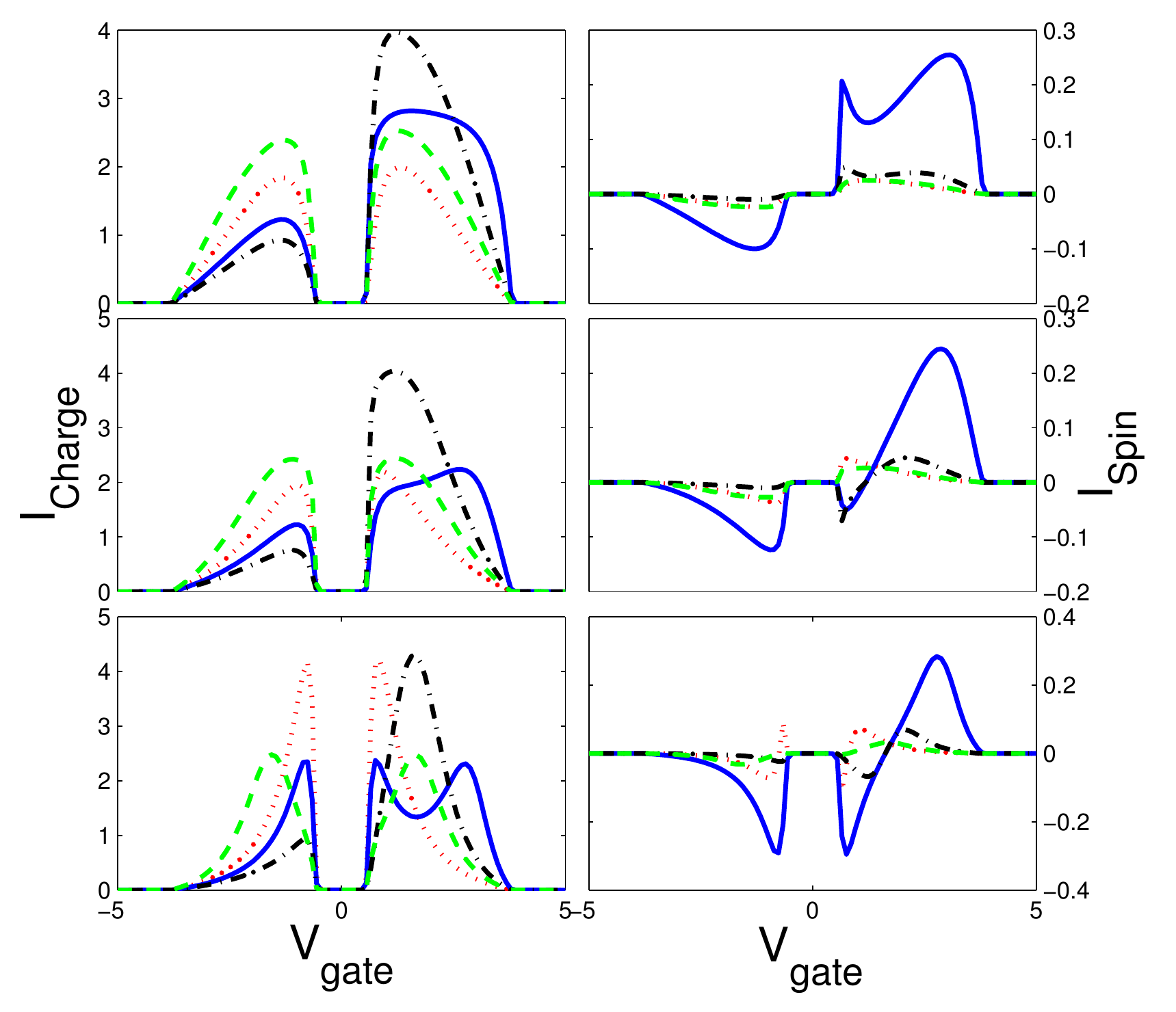}
\caption{The left and the right columns provide information about the charge and spin current versus gate voltage $V_g$ of the AB/QD/AB junction, for different parameters: (1)  $\epsilon=0.05$, $B_z=0.1$, $B_x=0.1$ (dotted red line); (2) $\epsilon=1.0$, $B_z=0.1$, $B_x=0.1$ (dashed-dotted black line); (3) $\epsilon=1.0$, $B_z=0.9$, $B_x=0.5$ (solid blue line); (4) $\epsilon=0.05$, $B_z=1.0$, $B_x=0.1$ (dashed green line), respectively. The top, the middle and the bottom panels correspond to the 0, 1 and 2 surface states, respectively.}
\label{fig4}
\end{figure}
%%%%%%%%%%%%%%%%%%%%%%%%%%%%%%%%%%%%%%%%
%%%%%%%%%%%%%%%%%%%%%%%%%%%%%%%%%%%%%%%%
\begin{figure}
 \includegraphics[width=1\columnwidth]{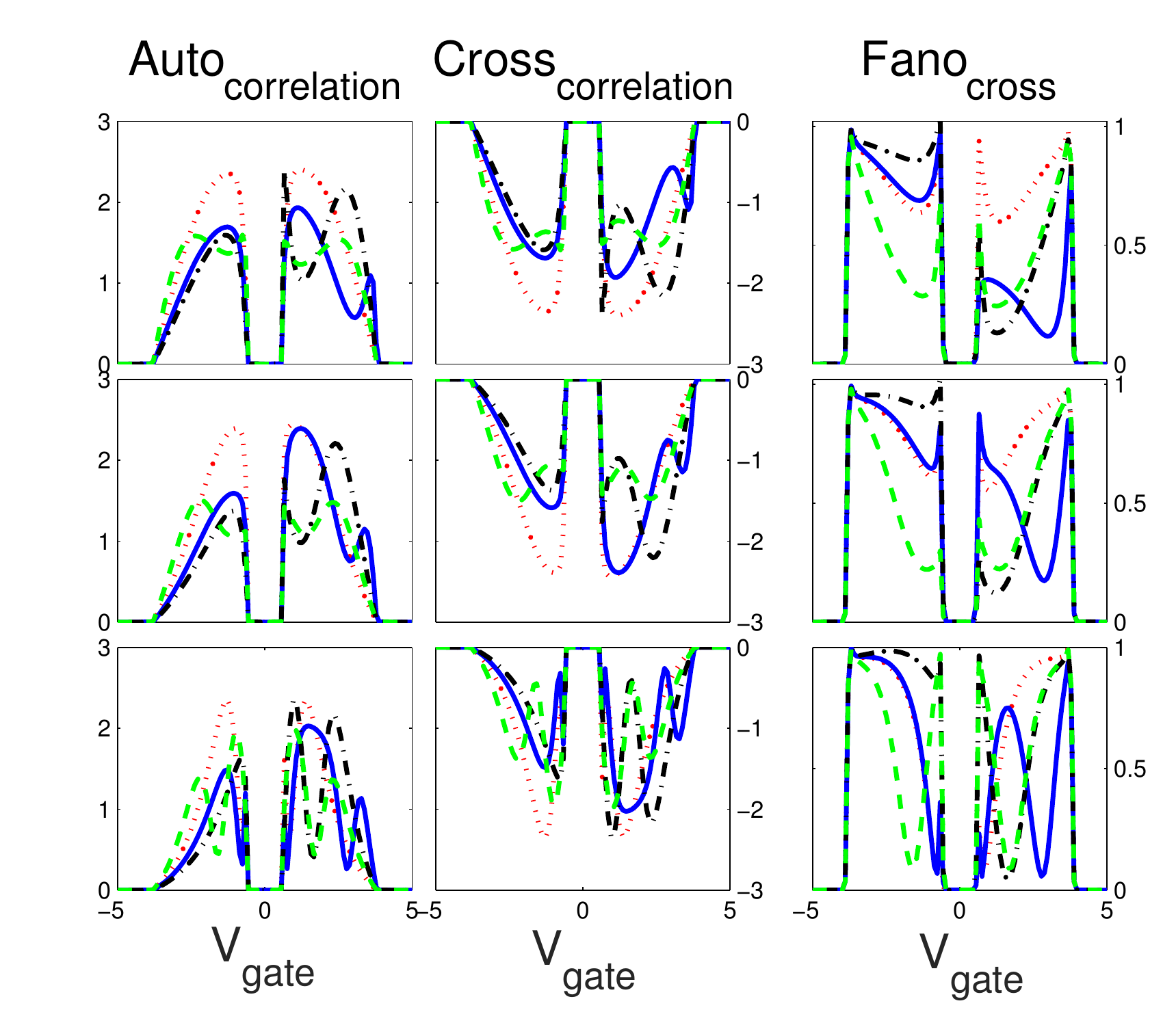}
\caption{The left, middle and the right columns indicate the charge auto-shot correlation , cross-shot correlation and cross-Fano factor versus gate voltage $V_g$ of the AB/QD/AB junction.  Parameters are same as the Fig.~\ref{fig4}.}
\label{fig5}
\end{figure}
%%%%%%%%%%%%%%%%%%%%%%%%%%%%%%%%%%%%%%%%
%%%%%%%%%%%%%%%%%%%%%%%%%%%%%%%%%%%%%%%%
\begin{figure}
 \includegraphics[width=1\columnwidth]{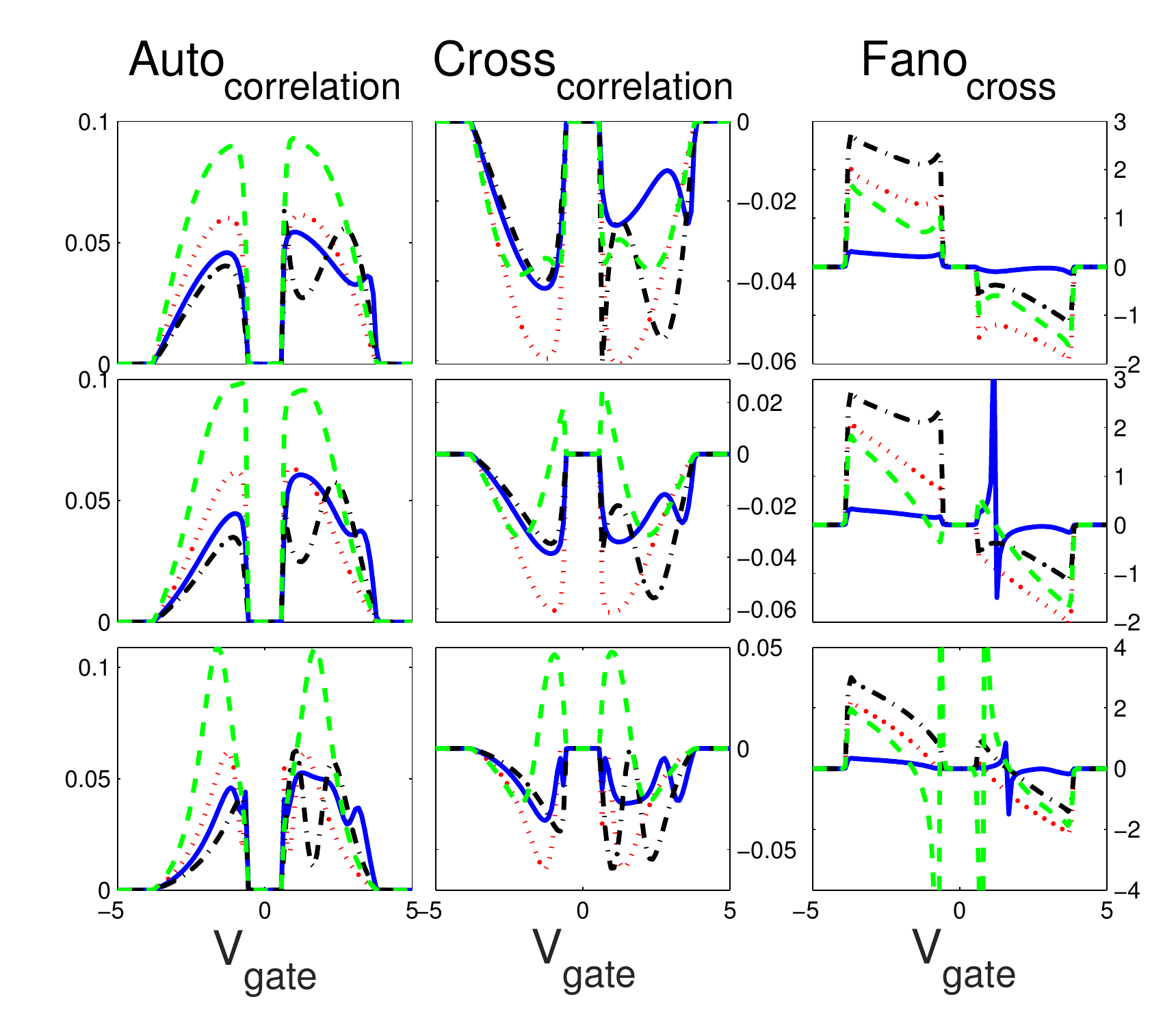}
\caption{The left, middle and the right columns indicate the spin auto-shot correlation , cross-shot correlation and cross-Fano factor versus gate voltage $V_g$ of the AB/QD/AB junction.  Parameters are same as the Fig.~\ref{fig4}.}
\label{fig6}
\end{figure}
%%%%%%%%%%%%%%%%%%%%%%%%%%%%%%%%%%%%%%%%
%%%%%%%%%%%%%%%%%%%%%%%%%%%%%%%%%%%%%%%%
\par
 For the AS case, when the QD is attached to the identical atom type on both leads (both $\varepsilon_k<0$ or $\varepsilon_k> 0$), the degenerate bonding configuration appear to create a system resonance at $\varepsilon_k$ for all site energy of te QD, which leads to perfect transmission at $E$ = $\varepsilon_k$ irrespective of the QD site energy $\varepsilon$, as shown in  Fig~\ref{fig3a}-(d). Conversely, when the QD has mixed bonding to the leads, the surface states at $\varepsilon_k$ and $-\varepsilon_k$ seem to destroy charge current at $\varepsilon_k$ (see Fig.~\ref{fig3a}-(b) and (f)). Many of the $HCRs^{,}$ trends observed in the charge current profile such as broadening, shifting and bending are still present for $HNSRs$ and $HPSRs$ for respective junctions (see Fig.~\ref{fig3b}-(b), (d) and (f)), The difference is that the spin currents change their sign in both bands for all surface states. Now, we proceed with the detailed study of each model separately.\\
\par
 \textbf{Alternative bond model(AB)}: In the left and right column of Fig.~\ref{fig4}, we have plotted the charge current and spin current versus gate voltage for different parameters, where top, middle and bottom panels correspond to 0, 1 and 2 surface states, respectively. Different line shapes  correspond to various parameters including  $\epsilon=0.05$, $B_z=0.1$, $B_x=0.1$ (dotted red line); $\epsilon=1.0$, $B_z=0.1$, $B_x=0.1$ (dashed-dotted black line); $\epsilon=1.0$, $B_z=0.9$, $B_x=0.5$ (solid blue line); $\epsilon=0.05$, $B_z=1.0$, $B_x=0.1$ (dashed green line), are shown. It can be seen from this figure, when the $\epsilon$ is so large compared to the value $\sqrt{B_{x}^2+B_{z}^2}$, the charge current has the lowest value in the negative range of gate voltage (dashed-dotted black line) by contrast, it has an opposite behaviour at the positive range of the gate voltage for all possible bonding configurations. As understood from the right column of this figure, for $\epsilon=1.0$, $B_z=0.9$, $B_x=0.5$ , the spin current (solid-blue line) has a more extensive range compared to the other three parameters for all possible surface states.  Another issue in the spin current profile is that, at some fixed parameters, different surface states show comlpicated line shapes in which one can see the two sign changing for the 2 surface state (see the dotted red line at the right bottom most panel of Fig.~\ref{fig4}). Moreover, there is an asymmetric behaviour of spin current between different signs of the gate voltage for other parameters containing $\varepsilon=1.0$  as we vary the gate voltage (it is more apparent for solid blue line). It is worth mentioning that for some fixed parameters ($\varepsilon=0.05$, $B_z=0.1$, $B_x=0.1$ (dotted red line)) one can see both charge and spin current enhancement in the 2 surface state (see the dotted red line at the  bottom  panels of Fig.~\ref{fig4}).
%  ================================\\
 %================================\\
 %================================\\
%  ================================\\
 %================================\\
\par
The left and middle  column of Fig.~\ref{fig5}(Fig.~\ref{fig6}), show the auto and cross of charge(spin) current noise versus gate voltage, where the top, middle and bottom panel are 0, 1 and 2 surface states, respectively.  The right column of these figures shows the cross-Fano factor of charge(spin) current. For charge current noise, both auto and cross behave as it is expected in which $S_{LL}=-S_{LR}$. A noticeable point which should be indicated is that at the edge of either valence  and conduction bands, the Fano factor reaches the Poissonian limit. Moreover, depends on parameters one can see symmetric or antisymmetric line shapes versus gate voltage.  In spin case (Fig.~\ref{fig6}), the auto correlation is positive for all possible surface states. When $\epsilon$ is so big compared to $\sqrt{B_{x}^2+B_{z}^2}$ (dashed-dotted black line) or is comparable, but smaller than this quantity(solid blue line), the auto correlation has a different behaviour in the positive and negative sign of the $V_g$. The former is followed by two peaks over the positive range. There is a symmetric behaviour of auto correlation between two different signs of the gate voltage for other parameters with $\epsilon=0.05$ (dotted red and dashed green line). Various line-shapes of correlation are found for different parameters. It is obvious that $S_{LL}\neq S_{LR}$. The middle column of this figure shows that the $S_{LR}$ has positive value over the gate voltage for 0 surface states as well as for all various parameters, while for 1 and 2 surface states it oscillates between negative and positive value for some parameters (dashed green line). The spin cross-Fano factor (see the right column of Fig.~\ref{fig6}.) shows quite an enhancement comparable to the charge one. For some parameters one can see the super-Poissonian limit reached. Similar to the metal case, in the AB junction, at the point which the spin current changes its sign (zero spin current),  a sharp enhancement of the spin Fano factor happens. Moreover, a huge enhancement also occurred  at the point which the auto correlation changes sign (see the dashed green line in the bottom panel).
\\
%%%%%%%%%%%%%%%%%%%%%%%%%%%%%%%%%%%%%%%%
%%%%%%%%%%%%%%%%%%%%%%%%%%%%%%%%%%%%%%%%
\begin{figure}
 \includegraphics[width=1\columnwidth]{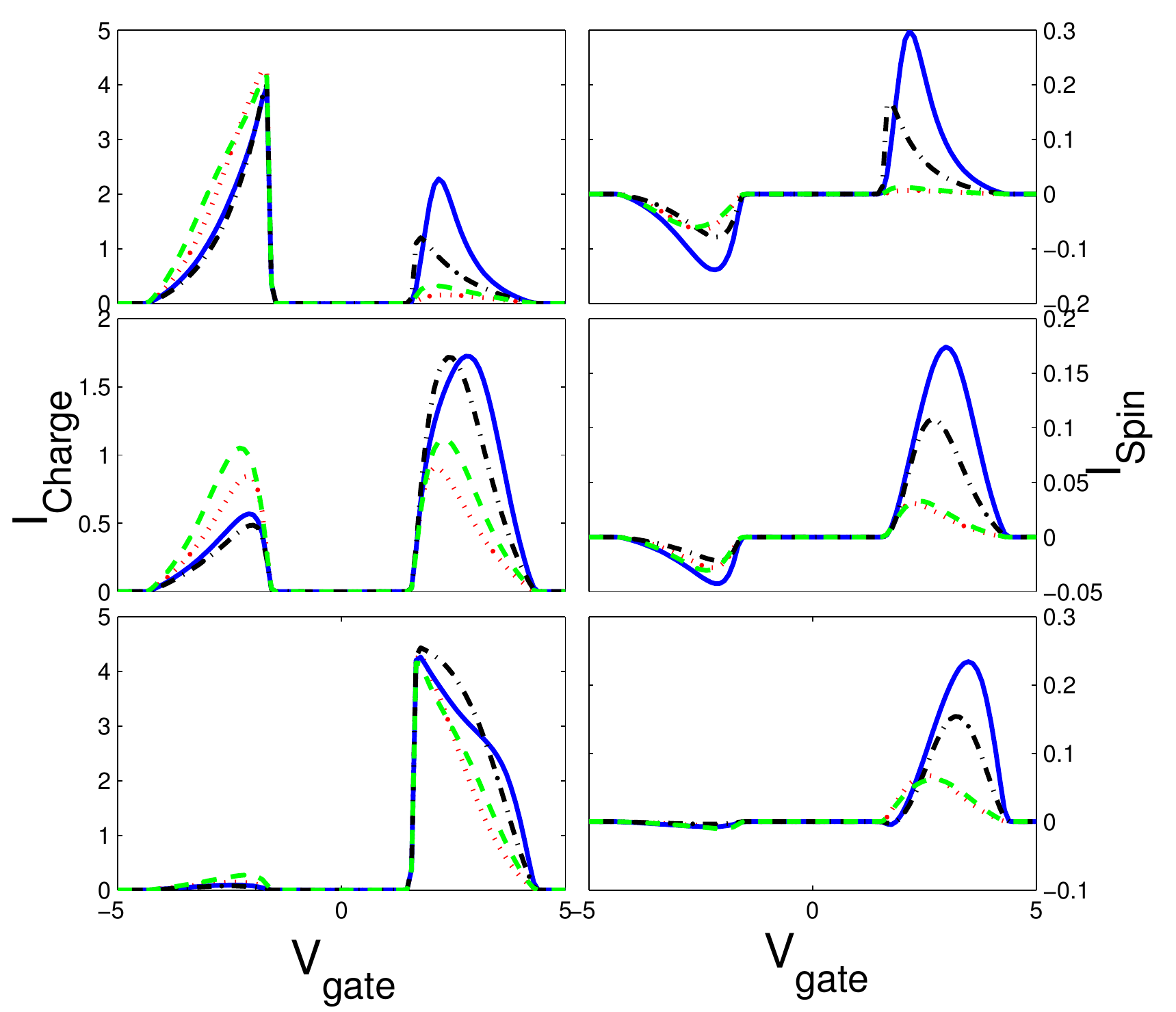}
\caption{The left and the right columns provide information about the charge and spin current versus gate voltage $V_g$ of the AB/QD/AB junction, for different parameters: (1)  $\epsilon=0.0$, $B_z=1$, $B_x=0.1$ (dotted red line); (2) $\epsilon=0.5$, $B_z=0.5$, $B_x=1.0$ (dashed black line); (3) $\epsilon=1.0$, $B_z=1.0$, $B_x=1.0$ (solid blue line), respectively. The top, the middle and the bottom panels correspond to the 0, 1 and 2 surface states, respectively.}
\label{fig7}
\end{figure}
%%%%%%%%%%%%%%%%%%%%%%%%%%%%%%%%%%%%%%%%
%%%%%%%%%%%%%%%%%%%%%%%%%%%%%%%%%%%%%%%%
\begin{figure}
 \includegraphics[width=1\columnwidth]{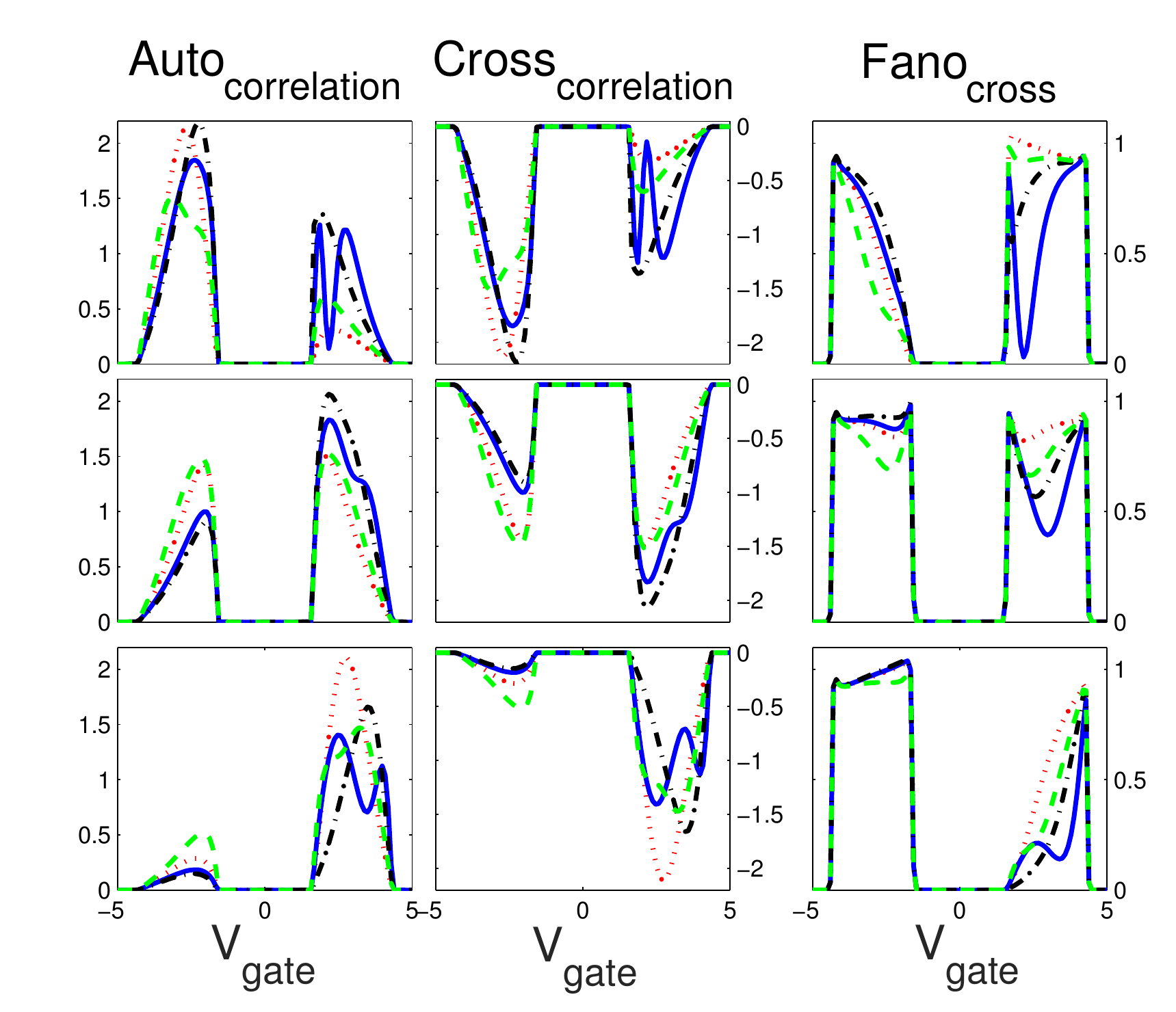}
\caption{The left, middle and the right columns indicate the charge auto-shot correlation , cross-shot correlation and auto-Fano  versus gate voltage $V_g$ of the AS/QD/AS junction.  Parameters are same as the Fig.~\ref{fig7}.}
\label{fig8}
\end{figure}
%%%%%%%%%%%%%%%%%%%%%%%%%%%%%%%%%%%%%%%%
%%%%%%%%%%%%%%%%%%%%%%%%%%%%%%%%%%%%%%%%
\\
\par
\begin{figure}
 \includegraphics[width=1\columnwidth]{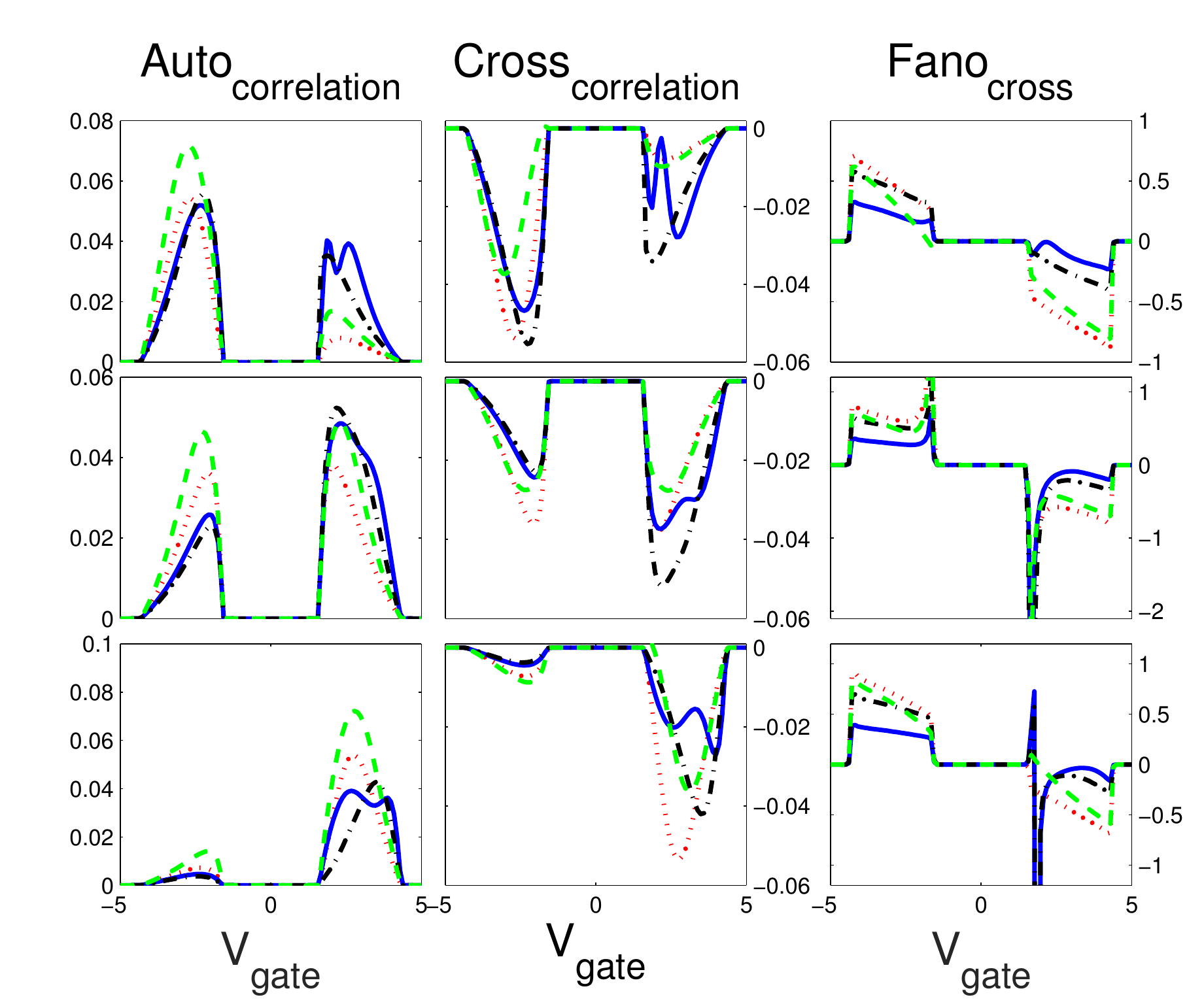}
\caption{The left, middle and the right columns indicate the charge auto-shot correlation , cross-shot correlation and auto-Fano  versus gate voltage $V_g$ of the AS/QD/AS junction.  Parameters are same as the Fig.~\ref{fig7}.}
\label{fig9}
\end{figure}

\textbf{Alternative site model(AS)}: Now, we regard a $TiO_2$/QD/$TiO_2$ connection. We face with similar choices in bonding configuration. In Fig.~\ref{fig7} we have depicted the results for charge current (left column) and spin current (right column) versus gate voltage, while the auto correlation and cross correlation spectrum of charge (spin) current  are given in the left column and middle column of Fig.~\ref{fig8}(Fig.~\ref{fig9}), respectively, which top, middle and bottom panel correspond to 0, 1 and 2 surface states. In the AS model charge current shows less symmetry than the AB model which is the direct consequence of density of states. The spin current profile shows that the general trend observed in the AB model is still present. For charge current noise, both auto and cross behave as it is expected. Similar to the AB model,  here at the edge of both valence  and conduction bands, the Fano factor reaches the Poissonian limit $F=1$. Further, depends on the parameters, symmetric line shapes versus gate voltage just observed in the 1 surface state.
\par
By comparing the results shown in Fig.~\ref{fig6} and Fig.~\ref{fig9}, we can see that the spin cross shot noise for different parameters has negative values for junctions with two titanium di-oxide electrodes, while, it oscillates between positive and negative values for junctions with two silicon electrodes for parameters including $\epsilon=0.05$, $B_x=0.9$ and $\epsilon=0.05$, $B_x=0.1$ in 1 and 2 surface states, respectively, as we vary the gate voltage. This is an interesting feature that the two different AB and AS junctions show different responses to the spin-flip proscess. Quite similar to the AB model, the spin cross-Fano factor (see the right column of Fig.~\ref{fig9}) shows an enhancement comparable to the charge one. For some parameters one can see the super-Poissonian limit reached. Similar to the metal case, in the AS junction, at the point which the spin current changes its sign (zero spin current),  a sharp enhancement of the spin Fano factor happens.
%%%%%%%%%%%%%%%%%%%%%%%%%%%%%%%%%%%%%%%%
%%%%%%%%%%%%%%%%%%%%%%%%%%%%%%%%%%%%%%%%
%%%%%%%%%%%%%%%%%%%%%%%%%%%%%%%%%%%%%%%%
%%%%%%%%%%%%%%%%%%%%%%%%%%%%%%%%%%%%%%%%
%############################################################
%######################                        ##########################
%######################      Section V        ##########################
%######################                        ##########################
%#############################################################
\section{Conclusion}
In this paper, using the scattering matrix approach, we have theoretically investigated the shot-noise of spin current through a quantum dot with a resonant level coupled with two metal or semiconductor electrodes, where two longitudinal ($B_x$) and transverse ($B_z$) magnetic fields are applied to the QD. We use two generalized tight-binding models, alternating bond (AB) and alternating site (AS) models in order to characterize the semiconductor contacts. The well known Newns-Anderson model has been also considered to characterize the metal contacts as a comparison benchmark. Because the symmetries are broken in the semiconductor contacts, surface states appear. Three different kinds of surface states labeled as 0, 1 and 2 have been regarded. We see that the spectral density of current fluctuations of spin current is not a conserved quantity. Hence both auto correlation and cross correlation are needed to characterize the correlation of spin current for a system including $B_x$. Model calculations show that, by tuning the gate voltage cross-correlation for different parameters has negative values for AS/QD/AS junctions, by contrast for AB/QD/AB junctions, it fluctuates between positive and negative values over the gate voltage for some parameters and surface states. This is an interesting feature that the two different AB and AS junctions has different response to the spin-flip proscess. Different line shapes are found as we change the system parameters in particular for semiconductor junctions.  We also addressed two kind of spin Fano factor correspond two auto and cross noise.  Results show that for both metal and semiconductor junctions, near the point which the spin current changes its sign (zero spin current) one can see a sharp enhancement of the Fano factor up to $F=2$ fro some parameters.  Actually,  it resembles  to the superconductor with $F=2$ which the Cooper pairing  causes an attractive Coulomb interaction between the electrons.\\

\section{ASSOCIATED CONTENT}
The Supporting Information is available free of charge via the internet at \url{http://pubs.acs.org}.
\par
In the supporting information, we have provided a detailed and pedagogical derivation of the Shot noise and self energies. In section 1, we introduce a graphical approach. Section 2 gives a full description of current-current fluctuation. Two equilibrium and non-equilibrium situations are explained in section 4 and section 5, respectively. A two terminal junction case is followed by section 5. In section 6, we present a lengthy, but straightforward calculation of self-energies.
%############################################################
%######################                        ##########################
%######################      Section VI       ##########################
%######################                        ##########################
%#############################################################
%\section{Acknowledgement}

%\section*{References}

\end{document}